\documentclass[12pt]{article} 
\usepackage{epsfig}
\usepackage{colordvi}
\def\bea{\begin{eqnarray}}
\def\eea{\end{eqnarray}}
\def\bq{\begin{quote}}
\def\eq{\end{quote}}

\font\tenrsfs=rsfs10 at 12pt
\font\sevenrsfs=rsfs7
\font\fiversfs=rsfs5
\newfam\rsfsfam
\textfont\rsfsfam=\tenrsfs
\scriptfont\rsfsfam=\sevenrsfs
\scriptscriptfont\rsfsfam=\fiversfs
\def\mathscr#1{{\fam\rsfsfam\relax#1}}
\def\Lag{\mathscr{L}}

\def\circa#1{\,\raise.3ex\hbox{$#1$\kern-.75em\lower1ex\hbox{$\sim$}}\,}

\makeatletter
\def\art{\@ifnextchar[{\eart}{\oart}}
\def\eart[#1]#2#3#4#5#6{{\rm #2}, {#3 \rm #4} {\rm (#6) #5} [{#1}]}
\def\hepart[#1]#2{{\rm #2, #1}}
\newcommand{\oart}[5]{{\rm #1}, {#2 \rm #3} {\rm (#5) #4}}
\makeatother

\newcommand{\NP}{Nucl. Phys.}

\newcommand{\PL}{Phys. Lett.}
\newcommand{\PR}{Phys. Rev.}

\def\gappeq{\mathrel{\rlap
{\raise.5ex\hbox{$>$}}
{\lower.5ex\hbox{$\sim$}}}}
\def\lappeq{\mathrel{\rlap{\raise.5ex\hbox{$<$}}
{\lower.5ex\hbox{$\sim$}}}}
\def\simlt{\stackrel{<}{{}_\sim}}
\def\simgt{\stackrel{>}{{}_\sim}}

\newcommand{\beq}{\begin{equation}}
\newcommand{\eeq}{\end{equation}}

\newcommand{\Un}[1]{\mb{V}_\nu^{#1}}

\def\tim{\tilde{m}_1}
\def\bmeg{\hbox{\rm BR}(\mu\to e\gamma)}
\def\bteg{\hbox{\rm BR}(\tau\to e\gamma)}
\def\btmg{\hbox{\rm BR}(\tau\to \mu\gamma)}

\def\mmaj{\mathbf{M}_R}

\def\varkappa{\omega}
\newcommand{\mb}[1]{\mbox{\normalsize\boldmath $#1$}}
\def\gev{\,\mathrm{GeV}}

\def\ldv{\vec{\lambda}_2}
\def\ltv{\vec{\lambda}_3}
\def\lth{\hat{\lambda}_3}
\def\ldh{\hat{\lambda}_2}
\def\ly{\mbox{\boldmath{$\lambda$}}}
\def\mnu{\mb{m}_\nu}
\def\mmaj{\mb{M}}
\def\gev{\,\mathrm{GeV}}

\newcounter{mnotecount}[section]

\begin{document}
\pagestyle{empty}
\begin{flushright}
{IFUP-TH/27-2004}\\
{IFT-2004/21}\\
{ECT*-04-09}\\
{hep-ph/0408015}\\
\end{flushright}
\vspace*{5mm}
\begin{center}

{\large {\bf\Large Low-scale standard  supersymmetric  leptogenesis }}\\
\vspace*{1cm}
{\bf Martti Raidal}$^1$,
{\bf Alessandro Strumia}$^2$  and {\bf Krzysztof Turzy{\'n}ski}$^{3}$\\
\vspace{0.3cm}

{\em
$^1$ National Institute of Chemical Physics and Biophysics,
Tallinn 10143, Estonia
\\
$^2$  Dipartimento di Fisica dell'Universit{\`a} di Pisa and INFN,  
Italia\\
$^3$  Institute of Theoretical Physics, Warsaw University, ul.~Ho\.za  
69, 00-681 Warsaw, Poland}

\vspace*{1.7cm}
{\bf\large Abstract}
\end{center}
\vspace*{1mm}
\noindent
\begin{quote}\large
Strictly adhering to the standard supersymmetric seesaw mechanism,
we present a neutrino mass model which allows successful standard  
thermal
leptogenesis compatible with gravitino cosmology. 
Some neutrino Yukawa couplings are naturally much
larger than the naive estimates
following from the seesaw formula.
This leads to large $\bmeg$,  detectable in
the next round of experiments. Ratios of $\mu\to e\gamma$,
$\tau\to e\gamma$ and $\tau\to\mu\gamma$ branching ratios are predicted  
in
terms of the measurable neutrino mass matrix.
\end{quote}
\vspace*{1.0cm}
\date{\today}


\vspace*{0.2cm}

\vfill\eject
\newpage

\setcounter{page}{1}
\pagestyle{plain}

\section{Introduction}

The observed neutrino masses can be explained by
adding to the Standard Model (SM) one heavy right-handed neutrino $N_i$
per generation~\cite{see-saw}:
\beq
\Lag = \Lag_\mathrm{SM} + \bar{N}_ii\partial\!\!\!/\,N_i + \left(  
\ly_{i\ell} N_i L_\ell H +\frac{1}{2} \mmaj_{ij} N_iN_j + \mathrm{h.c.}  
\right).
\eeq
The neutrino mass matrix depends on the unknown right-handed neutrino  
mass matrix
$\mmaj$ (with eigenvalues $M_1<M_2<M_3$) and neutrino Yukawa couplings  
$\ly$ as
\beq
\label{eqsee-saw}
\mnu=-v^2\ly^T\mmaj^{-1}\ly,
\eeq
suggesting that the masses of the right-handed
neutrinos are below $10^{14}\gev$. Unlike the SM, this extension can  
generate the
observed baryon asymmetry
\beq
\frac{n_B}{n_\gamma} = (6.15\pm 0.25) \times 10^{-10}
\eeq
via thermal leptogenesis in out-of-equilibrium decays of the lightest  
right-handed neutrino~\cite{fuya}.

Various considerations motivate supersymmetry broken by Fermi-scale
soft terms. Supersymmetric seesaw models have new signals: quantum  
effects
imprint the lepton flavour violation of the neutrino Yukawa couplings in
the left-handed slepton masses, giving rise to potentially observable
$\mu\to e\gamma$ and related lepton-flavour violating (LFV) processes.  
Their
rates depend on the unknown (but measurable) sparticle spectrum and,  
more
importantly, on the unknown neutrino Yukawa couplings $\ly$.
Since neutrino masses
tell us only $\ly^T\mmaj^{-1}\ly$, LFV rates cannot be predicted in  
terms of low-energy observables.

In addition, standard supersymmetric leptogenesis faces a potential  
problem:
gravitino decays destroy Big Bang Nucleosynthesis (BBN), unless the  
maximal `reheating'
temperature of the hot Big Bang $T_{\rm RH}$ is less than about  
$10^7\gev$
(the precise value depends on the gravitino mass and on its
hadronic branching ratio) \cite{gravitino}.\footnote{This bound does not
apply if gravitinos are the lightest stable SUSY
particles \cite{bobrbu}. However, in such a case, BBN can be destroyed  
by
decays of the
next-to-lightest SUSY particles \cite{gravitino}. Other scenarios of
leptogenesis that avoid the gravitino problem, e.g.~resonant  
leptogenesis
with almost degenerate singlet neutrino
masses~\cite{resonant1,resonant,ery1,piun,kt}, soft leptogenesis
\cite{soft,radia}, non-thermal leptogenesis \cite{nonth}, and  
leptogenesis
with $N_1$ or $\tilde{N}_1$-dominated universe  
\cite{hamuya,1April,gnrrs} have
been proposed. The gravitino problem is absent
in models where
the gravitino is ultraheavy or not present at all \cite{golung}.}
On the other hand, successful minimal leptogenesis
requires \cite{gnrrs,daib,bubapl}
\beq
\label{lgbound}
T_{\rm RH}>2\times 10^9\gev  ,
\eeq
if zero initial abundance of right-handed neutrinos is assumed and
only the leading $\mathcal{O}(M_1/M_{2,3})$ terms
are included in the calculation of the
CP asymmetry $\varepsilon_1$ in
decays $N_1\to LH, \bar L H^*$ of the lightest right-handed neutrino  
$N_1$.
These terms originate
from the same effective operator $(LH)^2$ which also gives
the contributions $\tilde{m}_{2,3}$ of the heavier $N_{2,3}$
to neutrino masses\footnote{In this limit the two
'vertex' and 'self-energy' loop diagrams collapse into a single diagram.
This observation could have allowed avoiding past debates (solved by
\cite{corovi})
on whether the 'self-energy' diagram contributes to $\varepsilon_1$.
See~\cite{hlnps} for a quantitative definition 
and discussion of $\tilde{m}_{2,3}$.}.
Thanks to this close relation, the naive estimate
$\varepsilon_1 \sim 3 \tilde{m}_{2,3}M_1/8\pi v^2$ can be improved to
  the rigorous Davidson-Ibarra bound ~\cite{daib,bcst}
\beq
\label{dibound}
|\varepsilon_1| \leq \varepsilon_1^\mathrm{DI} = \frac{3}{8\pi}  
\frac{(m_3-m_1)M_1}{v^2}
\eeq
where
$m_1 < m_2 < m_3$ are light neutrino masses.
Combined with precise studies of the dynamics
of thermal leptogenesis \cite{gnrrs}
this implies eq.~(\ref{lgbound}).

Higher orders corrections to this formula,
suppressed by powers of $M_1/M_{2,3}$, are not directly
related to neutrino masses and give the additional contribution
\beq
\label{ours}
\delta\varepsilon_1 \simeq \frac{3}{16\pi}  
\frac{\tilde{m}_{2,3}M_1}{v^2} \left( \frac{M_1}{M_{2,3}}\right)^2.
\eeq
A CP asymmetry much above the bound (\ref{dibound})
is obtained for $\tilde{m}_{2,3}/m_{2,3}\gg  
(M_{2,3}/M_1)^2$~\cite{hlnps}.
In other words, one needs that  the
$N_2$ and $N_3$ contributions to neutrino masses $\tilde{m}_{2,3}$
are orders of magnitude larger than the
neutrino masses $m_{2,3}$ themselves, i.e.~these contributions have to
cancel each other.
Barring anthrophic arguments and in absence of an underlying theoretical
justification this would be an unlikely fine-tuning.
While both papers in Ref.~\cite{hlnps} claim that it seems
possible to justify such cancellations,
the suggested pseudo-Dirac structure fails
because it controls neutrino masses, but
at the price of suppressing the CP-asymmetry of  
eq.~(\ref{ours})~\cite{hlnps}.

The goal of this paper consists in finding the class of
models which realizes the above mechanism naturally
and in studying their implications for the rates of LFV processes.
In section~\ref{model} we present the texture of the right-handed 
neutrinos mass matrix needed to control the appearance of the Yukawa
couplings in the light neutrino mass matrix so that these cancellations 
occur naturally, and outline how it can be justified by flavour models.
We also address the issue of the size of effective neutrino mass 
$\tilde m_1$ and show that it can be naturally small as required by
successful leptogenesis.
In section~\ref{fla} we perform a phenomenological study of such models
relating the model parameters to the neutrino oscillation measurements.
We show that the proposed class of models predicts
ratios of $\mu\to e\gamma$,
$\tau\to e\gamma$ and $\tau\to\mu\gamma$ branching ratios in
terms of the measurable neutrino mass matrix.
In section~\ref{lepto} we
show that such LFV rates must be sizable
in the region of the parameter space that allows
successful thermal leptogenesis
compatibly with the gravitino constraint.
In section~\ref{sub} we discuss
the stability of predictions with respect to small perturbations
from the `exact' texture.
Results are summarized in section 6.

\section{Building models with $|\varepsilon_1|\gg  
\varepsilon_1^\mathrm{DI}$}\label{model}
In this section
we outline the basic features of our neutrino mass model
which naturally accommodates
an enhanced CP-asymmetry $|\delta\varepsilon_1| \gg \varepsilon_1^{\rm  
DI}$.
For later convenience we introduce an unusual notation,
writing the neutrino Yukawa matrix as
\beq\label{nuYukawa}
\ly_{i\ell} = \left(\begin{array}{ccc} \lambda_1^e & \lambda_1^\mu &  
\lambda_1^\tau \\ \lambda_2^e & \lambda_2^\mu & \lambda_2^\tau \\  
\lambda_3^e & \lambda_3^\mu & \lambda_3^\tau \end{array} \right) =  
\left(\begin{array}{c} \vec{\lambda}_1 \\ \ldv \\ \ltv  
\end{array}\right)
\eeq
where $\vec{\lambda}_{1,2,3}$ are 3-vectors with $\ell=\{e,\mu,\tau\}$  
components.
We assume that the entries orders of magnitude
larger than what
na\"{\i}vely suggested by the seesaw 
(needed to get large $\tilde{m}_{2,3}$)
are all contained in  the Yukawa couplings $\vec{\lambda}_3$ of $N_3$.
For the moment we assume the simplest hypothesis suggested by data:
`lopsided' Yukawa couplings~\cite{baba},
i.e.~that each $\vec{\lambda}_i$ has
comparable entries $\lambda_i^{e,\mu,\tau}$, which naturally leads to  
large
mixing angles.\footnote{{\em A posteriori} in section~3 we will find
that for $\tilde{m}_1\ll m_{\rm sun}$ the model is predictive enough that 
our {\em a priori} assumption of lopsided $\vec\lambda_{2,3}$
is required by observed neutrino masses and mixings.}
Our assumption therefore reads
\beq\label{eq:3>>12} \lambda_3^\ell \gg \lambda_{1,2}^\ell  .\eeq
Finally, our key assumption is that, in the basis where eq.~(\ref{eq:3>>12}) holds,
the right-handed neutrino mass matrix has the following form
\beq
\label{texturem}
\begin{array}{cc} & \begin{array}{ccc} \,N_1\, & \,N_2\, & \,N_3\,  
\end{array} \\
\mmaj = \begin{array}{c} N_1 \\ N_2 \\ N_3 \end{array} \!\!\!\!\!&
\left( \begin{array}{ccc} M_{11} & 0 & 0 \\ 0 & 0 & M_{23} \\ 0 &  
M_{23} & M_{33}\end{array}\right)
\end{array} .
\eeq
The assumption $\mmaj_{12}=\mmaj_{22}=0$ implies vanishing 33 and 13
entries in the inverse mass matrix
\beq
\mmaj^{-1} = \left( \begin{array}{ccc} M_{11}^{-1} & 0 & 0 \\ 0 &  
-M_{33}/M_{23}^2 & M_{23}^{-1} \\ 0 & M_{23}^{-1} & 0  
\end{array}\right)
\eeq
that enters the seesaw formula
\beq
\label{see-sawf}
\mnu = -v^2 \sum_{i,j} (\mmaj^{-1})_{ij} \,\vec{\lambda}_i \otimes  
\vec{\lambda}_j.
\eeq
Therefore,
the large Yukawa couplings $\ltv$ do not generate large contributions
to the neutrino masses of  order  $\lambda_3^2v^2/M_{33}$. In the
right-handed neutrino mass basis, both $N_2^\mathrm{mass}$ and
$N_3^\mathrm{mass}$ give large contributions
$\tilde{m}_{2,3}\sim \lambda_3^2v^2/M_{2,3}$ to neutrino masses, 
but they naturally cancel  out.

The matrix $\mmaj^{-3}$, which controls the `higher order' contributions
to $\varepsilon_1$, has no special structure in the $N_2$, $N_3$ sector:
\beq
\mmaj^{-3} = \left( \begin{array}{ccc} M_{11}^{-3} & 0 & 0 \\ 0 &  
\mathcal{O}(M_{23}^{-3}) & \mathcal{O}(M_{23}^{-3}) \\ 0 &  
\mathcal{O}(M_{23}^{-3}) & -M_{33}/M_{23}^4 \end{array}\right)
\eeq
and its 33 entry gives the dominant contribution to $\varepsilon_1$,  
proportional
to $\lambda_3^2$.

A crucial issue for successful leptogenesis is 
the smallness of the effective neutrino mass $\tilde m_1$,
needed to have enough out-of-equilibrium $N_1$ decays.
The further assumption $\mmaj_{13}=0$ (together with  
$\mmaj_{12}=0$)
guarantees that the large Yukawa coupling $\lambda_3$ does
not contribute to $\tilde{m}_1$.
Indeed,  $N_1$ is already the lightest
mass-eigenstate which operates leptogenesis, with mass $M_1 = M_{11}$
and  $\tilde{m}_1 =  (\vec\lambda_1^*\cdot\vec\lambda_1) v^2/M_{11}$,
which is therefore small.

Thus the texture described between equations~(\ref{nuYukawa}) 
and (\ref{texturem}) 
has, in principle, the features which
allow successful leptogenesis
and reproducing the observed neutrino masses and mixings.
We shall discuss these issues more thoroughly in the following
sections.


So far we identified a special basis in which our initial problem gets reduced
to the simpler problem of explaining why certain entries are small.
We now discuss how the needed texture can be theoretically justified by building  
appropriate flavour models.
This second step usually does not lead to further insights, 
because usually there are many possible ways of justifying small couplings.
This will turn out to be true also in our case, which however presents a potential difficulty:
 the texture eq.~(\ref{texturem})  need strict  zeroes in specific entries.
 Therefore we here explicitly address only this potentially difficult issue,
briefly presenting one example
with $M_{23}\simgt M_{33}$ (models with $M_{23}\simlt M_{33}$ look more
complicated).
We introduce a U(1) flavour symmetry under which
$N_1$, $N_2$, $N_3$, $(L_{e,\mu,\tau}H)$ have charges $0,1,-1,1$,
respectively, and which is broken by the vev of a flavon superfield  
$\varphi_+$
with charge 1. This implies $\mmaj_{12}=\mmaj_{22}=0$,
$\vec{\lambda}_1=\ldv=0$ and allows a `lopsided' $\ltv$.
One can force $\mmaj_{13}=0$
by assuming that $\varphi_+$ has charge $2/3$ (rather than $1$),
or by adding extra symmetries that act only on $N_1$ or by assuming that
$N_{1}$ and $N_{2,3}$ have wave-functions localized in extra dimensions
  and with small overlap among them.
Finally, assuming that the flavon $\varphi_-$ with charge equal to  
$\varphi_+^\dagger$ has a small vev
$\langle\varphi_-\rangle \ll \langle\varphi_+\rangle$, which is natural
in supersymmetric models\footnote{The $F$-term potential, being  
invariant under the complexified global flavour symmetry,
determines at most the product $\langle \varphi_+ \varphi_-\rangle$.
The individual vevs might be fixed by SUSY-breaking corrections;
the desired hierarchy  $\langle\varphi_-\rangle \ll  
\langle\varphi_+\rangle$
can be naturally obtained e.g.\ when $\langle \varphi_+  
\varphi_-\rangle=0$ in the SUSY-limit.
`Supersymmetric zeroes' are discussed in~\cite{AF}.}, allows small  
`lopsided' $\vec{\lambda}_{1,2}$
and small $\mmaj_{12}$, $\mmaj_{22}$.
Note that the mechanism that ensures $\mmaj_{13}=0$ also breaks the  
relation $\lambda_1^2\sim \lambda_2$, expected from our minimal charge  
assignment.
One possibility of generating $\lambda_1$ and $\lambda_2$ in the
phenomenologically interesting range (discussed later) is the  
introduction of
four flavons with charges $\pm2/3$ and $\pm1$ and with appropriate  
vevs\footnote{We thank
S.~Davidson for suggesting models with such kinds of charge assignments  
and for indicating us the features we presented.}.
The U(1) flavour symmetry can be extended
to the quarks and right-handed leptons in a SU(5)-invariant way,  
reproducing
the masses and mixings of all SM fermions.
It should be clear that proceeding as outlined above we could build many full models. 
We here prefer to study the general testable consequences.

Small entries of $\mmaj$ and $\ly$ are stable under quantum corrections.
This can be easily seen in SUSY models, where
quantum
corrections only renormalize kinetic terms. Furthermore, at tree level,
$N_{1,2,3}$ in general already have a non-canonical kinetic matrix.
Since it does not affect the seesaw mechanism we can ignore it.


\section{Flavour violation}\label{fla}

In this section we perform a thorough phenomenological study of the
proposed scenario with the aim to relate model parameters to the
neutrino measurements.   
We start with studying the  values of the neutrino Yukawa couplings  
$\vec{\lambda}_{1,2,3}$
with emphasis on the entries of the largest Yukawa couplings
$\ltv$, which are the
dominant source of the LFV effects. So far we assumed that each  
$\lambda_i$ is
roughly equally distributed among  flavours. We shall discuss  
carefully, how much this assumption is required by the observed
large atmospheric and solar mixing angles,
obtaining precise predictions for the flavour content of the neutrino  
Yukawa
couplings.

We choose
$\tim\sim 10^{-3}\,\mathrm{eV}$, which yields
\beq
\label{lambda1}
\lambda_1 \equiv (\vec{\lambda}_1^*\cdot \vec{\lambda}_1)^{1/2} =  
\frac{\sqrt{\tim M_1}}{v} \sim 10^{-6} \frac{M_1}{10^6\gev} .
\eeq
Together with large $\lambda_3$,
these choices make $n_B$ as large as possible\footnote{This might be  
however
unnecessary, allowing smaller or larger values of $\lambda_1$ and  
smaller
values of $\lambda_3$. We will discuss this issue later.}.
Then the neutrino mass matrix is dominated by the term
\beq
\label{mnuansatz}
\mnu = -\frac{v^2}{M_{23}} \left( \ltv\otimes\ldv + \ldv\otimes\ltv  
\right) ,
\eeq
which can give both the atmospheric and solar masses $m_2 =  
\sqrt{\Delta m^2_{\rm sun}}$,
$m_3=\sqrt{\Delta m^2_{\rm atm}}$
for
\beq
\label{lambda2}
\lambda_2 \sim \frac{m_{2,3}M_{2,3}}{\lambda_3v^2} \sim 10^{-8}  
\frac{1}{\lambda_3} \frac{M_{2,3}}{10^7\gev} .
\eeq
We now invert the relation~(\ref{mnuansatz}) between  seesaw  
parameters
and observable neutrino masses and mixings.
The analytic computation will be performed in an unusual but convenient  
way:
as suggested by the vector notation $\vec\lambda_i$
we will use vector calculus.\footnote{All our subsequent  
expressions
can be converted into traditional notations by replacing vectors with  
their components.
Models with neutrino masses and/or leptogenesis dominated
by two right-handed neutrinos have been studied previously in the  
literature,
but only performing generic analyses~\cite{GEN,ceprt} or
focusing on specific models~\cite{SPE} different from the ones under  
consideration.

The computation below
could be alternatively performed in a more indirect way using the $R$-matrix introduced  
in~\cite{see-sawparam},
after finding the specific class of $R$ matrices corresponding to our  
class of models.
Such $R$ matrices can be parametrized as
$$ R \simeq \pmatrix{1 &0 &0 \cr 0 & z & iz\cr 0 & -iz & z}\qquad  
\hbox{for $|z|\gg 1$}.$$
The phase of the complex number $z$ does not play any r{o}le.
See also~\cite{hlnps,ceprt}.  In particular,
our predictions for LFV rates are equivalent to appropriate limits of
equations~(18), (19) in~\cite{ceprt}.}

Computing neutrino masses $m_i\geq 0$ and neutrino
mass eigenstates $\hat\nu_i$ amounts to decompose the neutrino mass matrix
given in eq.~(\ref{mnuansatz}) as
\beq
\mnu = \sum_{i=1}^3 m_i \,\hat{\nu}^*_i \otimes \hat{\nu}^*_i
\eeq
where $\hat{\nu}_i$ are 
orthonormal vectors spanning the flavour space of 
the left-handed doublets. 
In the basis of  charged lepton mass eigenstates they are usually parametrized as
\beq
\hat{\nu}_3^\ell = \Un{\ell 3} \qquad \hat{\nu}_2^\ell = \Un{\ell  
2}e^{i\alpha} \qquad \hat{\nu}_1^\ell = \Un{\ell 1}e^{i\beta}
\eeq
where $\Un{}$ is the CKM-like part of the neutrino mixing matrix and
$\alpha$, $\beta$ are Majorana phases.

Defining the moduli $\lambda_i \equiv (\vec{\lambda}_i^*\cdot  
\vec{\lambda}_i)^{1/2}$,
the unit versors
$\hat{\lambda}_i\equiv \vec{\lambda}_i/\lambda_i$
and $\vec{\lambda}_3^*\cdot \vec{\lambda}_2\equiv \lambda_2\lambda_3  
ce^{2i\psi}$,
where $0\leq c\leq 1$ is the `cosine' of the angle between $\ldv$ and  
$\ltv$,
we get:
\beq\label{m123}
m_1=0 ,\qquad m_{2,3} = \lambda_2\lambda_3 (1\mp c) v^2/M_{23}.
\eeq
The observed moderate mass ratio $m_3/m_2\sim 5$ between the  
atmospheric and
solar masses is reproduced for $c\approx\cos 45^o$, which
does not require any special alignment of $\hat{\lambda}_2$ and
$\hat{\lambda}_3$ in the flavour space,
consistently with our assumption of
`lopsided' vectors $\vec{\lambda}_{2,3}$.
In contrast, the inverted hierarchy of the neutrino masses
requires $c \simeq \Delta m^2_{\rm sun}/4\Delta m^2_{\rm atm}\sim  
10^{-2}$,
which means that $\hat{\lambda}_2$ and $\hat{\lambda}_3$ have to
be almost orthogonal in the flavour space. Such an alignment
would clearly be less natural and need further
justification.\footnote{If nevertheless neutrinos turn out to have  
an inverted spectrum,
our predictions for this case can be found by substituting
$V^{\ell 3}_\nu$ with $V^{\ell 1}_\nu$ in eq.~(\ref{venti}).}

The neutrino mass eigenstates are
\beq\label{eq:19}
\hat{\nu}_1 \propto \lth \times \ldh \qquad \hat{\nu}_2 =  
\frac{-e^{i\psi}\lth^*+e^{-i\psi}\ldh^*}{\sqrt{2(1-c)}} \qquad  
\hat{\nu}_3 = i\frac{e^{i\psi}\lth^*+e^{-i\psi}\ldh^*}{\sqrt{2(1+c)}} .
\eeq
Inverting this relation we get
\beq
\hat{\lambda}_{3,2} = e^{\pm i(\psi\pm\pi/2)} \frac{\hat{\nu}^*_3\pm i  
\sqrt{m_2/m_3}\hat{\nu}^*_2}{\sqrt{1+m_2/m_3}}
\eeq
in terms of the low energy observables (not affected by the phase  
$\psi$).
Coming back to the traditional notation the flavour content of the
big neutrino Yukawa coupling is
\beq
\label{venti}
\lth^\ell = ie^{i\psi} \frac{\Un{\ell 3} + ie^{i\alpha}\Un{\ell 2}  
\sqrt{m_2/m_3}}{\sqrt{1+m_2/m_3}}.
\eeq
Changing $\alpha$ by $\pi$ is equivalent to swapping $\lth$ and $\ldh$.
In view of observed  large neutrino mixing angles,
the Yukawa coupling vectors $\hat{\lambda}_2^\ell$ and  
$\hat{\lambda}_3^\ell$
turn out to generically have `lopsided'  form,
which  justifies our heuristic assumption made in Section  
\ref{model}.\footnote{Only
if future experiments establish
$\theta_{13}$ much below present bounds, eq.~(\ref{eq:19}) will need  
unnatural cancellations indicating the necessity of a flavour symmetry  
more complicated than `lopsided'.
Our predictions would remain unchanged, as they follow from the   
texture in eq.~(\ref{texturem}),
not from the flavour models that can justify it.}

\begin{figure}[t]
\begin{center}
\begin{tabular}{ccc}
$\ref{figA}a$ && $\ref{figA}b$ \\
\includegraphics[height=7cm]{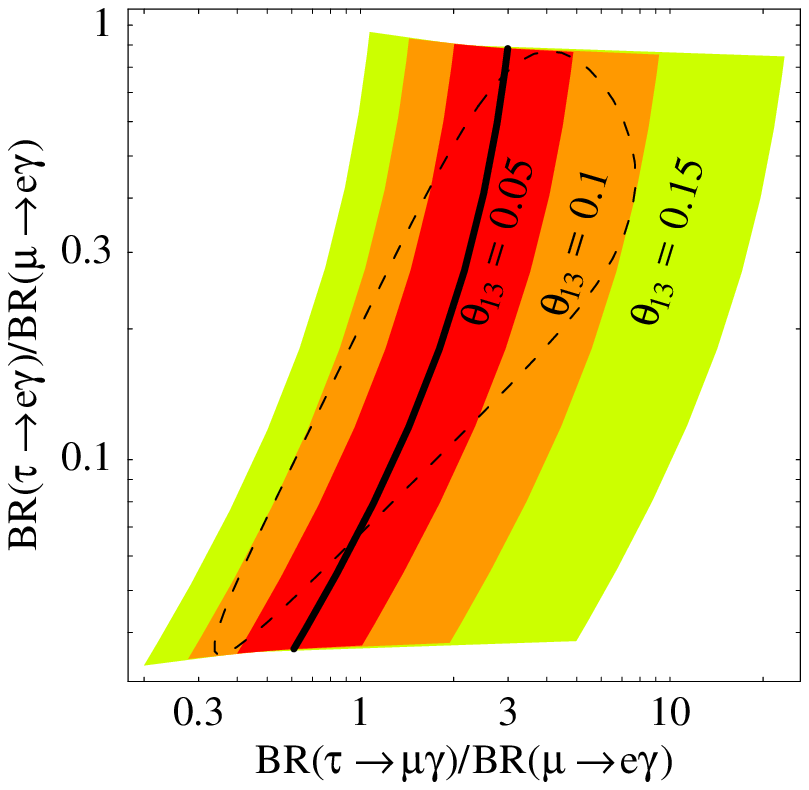} & \qquad &
\includegraphics[height=7cm]{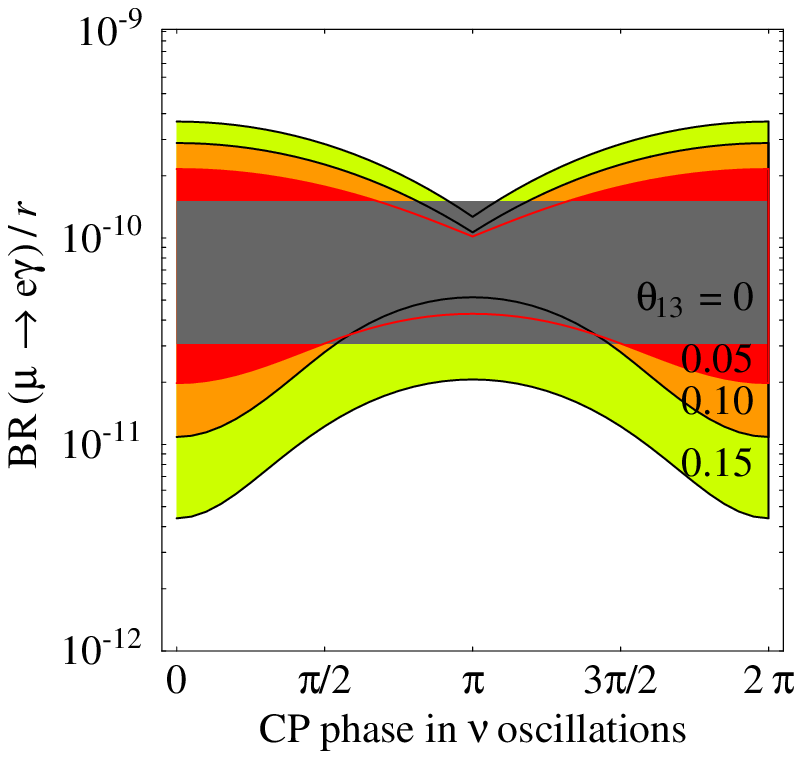}
\end{tabular}
\end{center}
\caption[X]{\em a)  Ratios of LFV decays allowed for  $\theta_{13} =
0$ (thick black line), $0.05, 0.1, 0.15$ for arbitrary
values of the Majorana phase  $\alpha$ and of the CP-
phase  $\phi$ in neutrino oscillations. The dashed line
shows the residual allowed range assuming that
future measurements will fix  $\theta_{13}=0$ and $\phi=1$.
b) Range of $\bmeg$ for arbitrary values of the Majorana phase
$\alpha$ as a function of the CP Dirac phase $\phi$ for different values
of $\theta_{13}$.
The factor $r\approx (\tan\beta/5)^2(200\gev/M_\mathrm{SUSY})^4$
equals one at our reference SUSY spectrum.
We fixed $\lambda_3 = 0.1$, $M_1 = 10^6\gev$, $M_3/M_2=2$.
  \label{figA}}
\end{figure}

It follows from eq.~(\ref{venti}) that the model predicts
\beq
\frac{\btmg}{\bmeg} = b\frac{|\lambda_3^\tau|^2}{|\lambda_3^e|^2},  
\qquad
\frac{\bteg}{\bmeg} = b\frac{|\lambda_3^\tau|^2}{|\lambda_3^\mu|^2}
\eeq
(where $b=BR(\tau\to\mu\bar{\nu}_\mu\nu_\tau)=0.18$) in terms of the
neutrino masses, mixings and CP phases measurable at low energy. We fix  
the
already measured quantities to their present best-fit values
\beq
\theta_{23}=\frac{\pi}{4}, \qquad \tan^2\theta_{12}=0.4, \qquad  
\frac{m_3}{m_2}=5,
\eeq
and show in  fig.~\ref{figA}a the predicted range of ratios among  
different branching
ratios. We  fix $\theta_{13}$ at a few possible values and freely vary
the remaining unmeasured parameters, i.e.~the Majorana phase $\alpha$  
and the
CP-phase $\phi$ in neutrino oscillations. The dashed line is an example
of the predicted range if all parameters except $\alpha$ were
measured. The model predicts that $\mu\to e\gamma$, $\tau\to e\gamma$  
and
$\tau\to\mu\gamma$ have comparable branching ratios. Since $\bmeg$ is
presently the most strongly constrained, only $\mu\to e\gamma$ seems  
observable
in the next round of experiments.

Note that it is not possible to have $\lambda_3^e=0$ and therefore
negligibly small $\bmeg$ and $\bteg$, since this would require
\beq\label{canc}
\tan\theta_{13}= \sqrt{m_2/m_3} \sin\theta_{12}
\eeq
which violates the `{\sc Chooz}'~\cite{CHOOZ} constraint on $\theta_{13}$.
More precisely, taking into account that $\theta_{13}$ is also constrained
by `solar' and `atmospheric' experiments, by performing a global fit we find
$\sin^2\theta_{13} = 0.007\pm 0.016$: data disfavour the relation~(\ref{canc})
at about $3\sigma$.

\begin{figure}[t]
\begin{center}
\begin{tabular}{ccc}
$\ref{fig2}a$ & &$\ref{fig2}b$ \\
\includegraphics*[height=7cm]{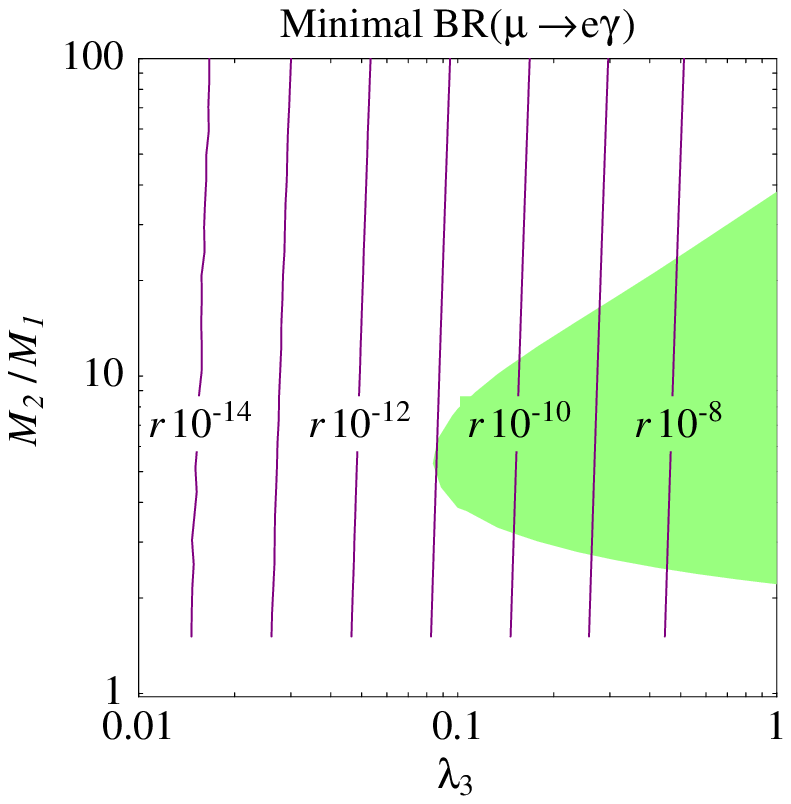} & \qquad&
\includegraphics*[height=7cm]{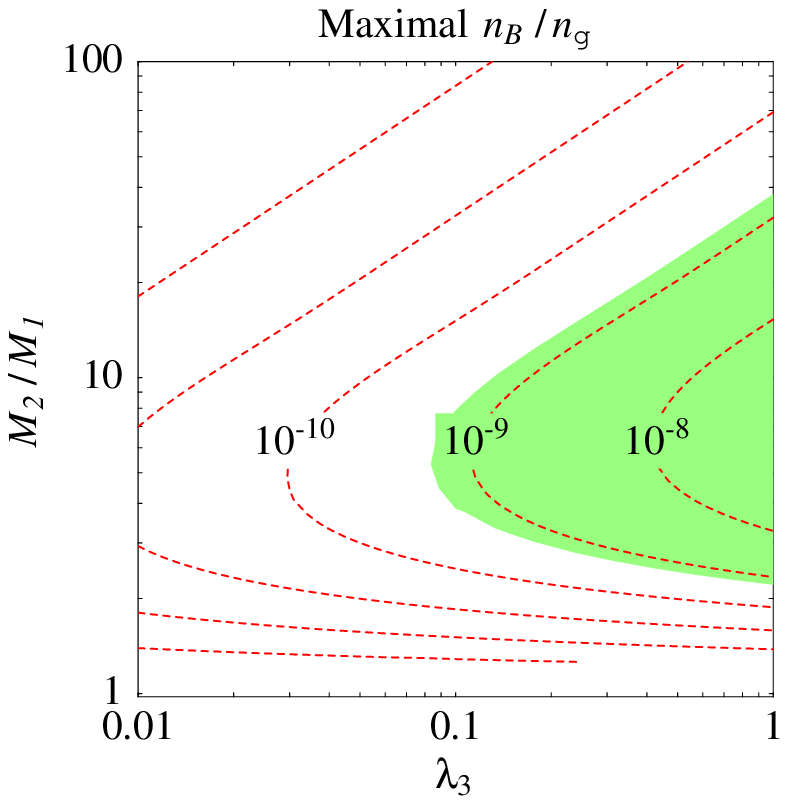}
\end{tabular}
\end{center}
\caption[X]{\em
As function of $\lambda_3$ and of $M_2/M_1$ we plot contour lines of:
(a) the minimal value of $\bmeg$;
the factor $r\approx (\tan\beta/5)^2(200\gev/M_\mathrm{SUSY})^4$
equals one at our reference SUSY spectrum.
(b) the maximal baryon asymmetry $n_B/n_\gamma$.
  The shaded region covers the allowed region.
  We assumed $\theta_{13}=0.1$, $\phi=1$,
  $M_1 = 10^6\gev$, $M_3/M_2 = 2$, $\tilde{m}_1 = 10^{-3}\,\hbox{\rm  
eV}$,
  but the main conclusions do not depend on these choices,
  as explained in the text.
  \label{fig2}}
\end{figure}

As usual, the predicted LFV rates depend on sparticle masses which can
be measured at colliders. Taking into account naturalness considerations
and experimental bounds and hints, we give our numerical examples for
$m_0=100\gev$, $M_{1/2}=200\gev$, $A_0=0$, $\tan\beta=5$.
$\bmeg$ is then approximately given by
\beq
\label{bmegest}
\bmeg \approx 2\cdot 10^{-7}r |\mb{Y}_{e\mu}|^2 \qquad \textrm{  
where}\qquad \mb{Y}=\ly^\dagger  
\ln\left(\frac{M_\mathrm{GUT}^2}{\mmaj\mmaj^\dagger}\right)\ly
\eeq
and $r\approx (\tan\beta/5)^2(200\gev/M_\mathrm{SUSY})^4$
  equals 1 at our reference point.\footnote{See, however, \cite{ceprt}  
for a discussion of potentially sizable corrections to this simple  
estimate.
  Furthermore, if $M_1$ is lowered down to Fermi-scale energies
  tree-level exchange of right-handed neutrinos
  gives  extra contributions to LFV rates
  beyond the supersymmetric loop contributions
  discussed here.
  }
Figures~\ref{figA} and \ref{fig2} show results of a precise computation,
performed by solving numerically the full set of the
renormalization group equations of the supersymmetric seesaw model
(see e.g.~\cite{kt,chapo})
and using exact formulae of \cite{hino}.

In fig.~\ref{figA}b we show how $\bmeg$
depends on $\theta_{13},\phi$ and $\alpha$
at fixed values of high-energy parameters.
In fig.~\ref{fig2}a we show how  $\bmeg$
(for $r=1$, $\theta_{13}=0.1$, $\phi=1$ and
minimized with respect to Majorana phases)
depends on the main
high-energy parameter, $\lambda_3$.
As illustrated in the figure and discussed in the next section,
successful thermal leptogenesis
demands a value of $\lambda_3$ that implies
$\mu\to e\gamma$ at a level observable
in the next round of experiments
(unless sparticles are much above the Fermi scale and
unless SUSY breaking is mediated to sleptons at energies below $M_1$).

\section{Leptogenesis}\label{lepto}

Gravitino cosmology probably requires $M_1$ a few orders of magnitude  
below
its minimal value $M^\mathrm{DI}_1 \sim 2\times 10^9\gev$ allowed by the
bound (\ref{dibound}) on the CP asymmetry  
$\varepsilon_1$~\cite{gravitino,daib}.
In our model this can be achieved thanks to a $\varepsilon_1$ orders of
magnitude larger than $\varepsilon_1^\mathrm{DI}$:
\beq\label{eq:eps1}
|\varepsilon_1|\le
  \varepsilon_1^\mathrm{DI}+
   \frac{3}{16\pi}\frac{(M_3-M_2)M_1^3}{M_{2}^2M_3^2}\lambda_3^2.
\eeq
(In the limit $M_3\gg M_2$ the maximal CP-asymmetry reduces to the  
estimate of~\cite{hlnps}).
The CP asymmetry needed to reach a given $M_1<M_1^\mathrm{DI}$ is
$|\varepsilon_1|=\varepsilon_1^\mathrm{DI}M_1/M_1^\mathrm{DI}$ (in the  
absence
of additional washout effects) and is obtained for
\beq
\label{l3est}
\lambda_3 \approx \frac{\sqrt{m_3 M_1^\mathrm{DI}}}{v} \left(  
\frac{M_{2,3}}{M_1}\right)^{3/2} \approx  
0.1\,\left(\frac{M_{2,3}}{10\,M_1}\right)^{3/2},
\eeq
where we approximated $M_{2,3}\approx M_2\approx M_3$.
Relatively large values of $\lambda_3$ are needed irrespectively of the
value of $M_1/M_1^\mathrm{DI}$, since for smaller $M_1$ the
asymmetry $\varepsilon_1$ needs more enhancement by the Yukawa  
couplings,
whose na\"{\i}ve values become smaller.
The CP asymmetry is saturated for $|{\rm Im}\,\hat{\lambda}_1\cdot  
\hat{\lambda}_3^*|=1$,
which can be achieved compatibly
with all experimental constraints on neutrino masses and mixings.
This becomes particularly clear for $\tilde{m}_1\ll m_{\rm sun}$,  
because in this case $\vec\lambda_1$
negligibly contributes to the neutrino mass matrix,
so that $\hat\lambda_1$ remains fully undetermined.
`Lopsided' Yukawa couplings suggest the generic expectation
$|{\rm Im}\,\hat{\lambda}_1\cdot \hat{\lambda}_3^*|\circa{<}1$.


The ratio $M_{2,3}/M_1$, dictating the size of $\lambda_3$, is further
constrained by leptogenesis: since $N_{2,3}$ have anomalously large  
Yukawa
couplings, on-shell $N_{2,3}$ exchange gives a significant extra
washout of $n_B$, unless $M_{2,3}/M_1$ is large enough that these
effects are strongly Boltzmann-suppressed during $N_1$ decays at $T\sim  
M_1$.
Roughly, one needs $e^{-M_{2,3}/M_1}\simlt  
\tilde{m}_{1}/\tilde{m}_{2,3}$ i.e.\
$M_{2,3}\simgt 10 M_1$.

Adopting the computation of \cite{gnrrs},
this issue is precisely studied in fig.~\ref{fig2}b, where we show the  
maximal
$n_B/n_\gamma$ as a function of $\lambda_3$ and of $M_2/M_1$ for  
$M_1=10^6\gev$,
$\tim=10^{-3}\,\mathrm{eV}$ and $M_3/M_2=2$
(which maximizes $|\varepsilon_1|$, see eq.~(\ref{eq:eps1})).
We  find that for $M_{2,3}/M_1 > 10$ the extra washout effects are  
negligible,
giving $\lambda_3$ in agreement with the estimate (\ref{l3est}).
At smaller $M_{2,3}/M_1$ the CP asymmetry $\varepsilon_1$ becomes  
larger, but the
baryon asymmetry $n_B$ decreases due to $N_{2,3}$-induced washout.
One needs $\lambda_3 > 0.1$, which implies $\bmeg$ close to present
experimental limits. This is a robust conclusion, not limited to the
specific values of $M_1$ and of $\tim$ chosen in  fig.~\ref{fig2}.  
Indeed the above
estimates dictate how  fig.~\ref{fig2} can be rescaled to other values  
of $M_1$ and
$\tim$ (as long as $N_1$ does not decay far from thermal equilibrium).
The precise value of the reheating temperature does not affect $n_B$,  
as long
as it is a few times larger than $M_1$. If $T_{\rm RH} > M_{2,3}$, also
$N_{2,3}$ decays contribute to leptogenesis, but without giving  
significant
corrections since they have negligible CP-asymmetries and big washouts.

\section{Effects of sub-leading terms}\label{sub}
In the previous sections we obtained predictions under the  
assumption that $N_3$ has large Yukawa couplings, see eq.~(\ref{eq:3>>12}),
which do not give rise to too large neutrino masses thanks
to exact  texture zeroes, see eq.~(\ref{texturem}).
This patten allows a natural
implementation of the mechanism proposed in~\cite{hlnps}.
A more generic class of natural models is obtained by
replacing the texture zeroes with generic entries, as long as they
are sufficiently small.
Using our vector notation we can now show,
 without any  new computation, that our predictions remain valid also in this case.

A large Yukawa coupling $\lambda_3$ is generically needed to lower  
$T_{\rm RH}$
by enhancing $\varepsilon_1$. Its minimal value can be better  
determined once
sparticles are discovered and their masses measured and understood.
The basic assumption that allowed us to control the flavour content
$\lth$ was that the 32 entry of $\mmaj^{-1}$
gives the dominant contribution to seesaw-induced neutrino masses.
This assumption followed from a model that naturally accommodates
a Yukawa coupling much larger than what is na\"{\i}vely
suggested by the seesaw.

The 33 and 31 entries of $\mmaj^{-1}$
would give extra contributions involving the big Yukawa coupling  
$\lambda_3$,
if the assumed texture zeros were not exact.
But even allowing these extra terms, the neutrino mass matrix can still
be written in the form of eq.~(\ref{mnuansatz}), with $\vec{\lambda}_2$
replaced by a linear combination of $\vec{\lambda}_{1,2,3}$ and with
$\vec{\lambda}_3$ unchanged.
Therefore, including these extra terms does not
affect the prediction for $\lth$ of eq.~(\ref{venti}) and the consequent
prediction for LFV rates.

The 11 and 12 entries of $\mmaj^{-1}$
do not give contributions to neutrino masses which involve the large  
$\ltv$
Yukawa coupling, and therefore can be neglected as long as $\lambda_1$  
and
$\lambda_2$ are sufficiently small. The smallness of $\lambda_2$ is  
demanded
by eq.~(\ref{lambda2}) (barring  fine-tunings), while the smallness of
$\lambda_1$ is only suggested by leptogenesis, eq.~(\ref{lambda1}).
Predictions for LFV rates can be changed at the expense of a larger
$\tim \simgt m_\mathrm{sun}$, which makes leptogenesis less efficient.  
This can
be compensated by a larger $\varepsilon_1$ obtained from a larger  
$\lambda_3$.
In this way (and without caring of the naturalness of the model)
it is also possible to suppress LFV effects by
concentrating  the big Yukawa coupling
$\vec\lambda_3$  into a single flavour
$\ell = e,\mu$ or $\tau$.
The big Yukawa coupling $\lambda_3^\ell$ still gives detectable  
effects, by making
$\tilde{\ell}_R$ lighter than the two other right-handed sleptons.

\section{Conclusions}\label{concl}
As pointed out in~\cite{hlnps}, the  DI  
bound~\cite{daib,bcst} can be evaded, allowing to reconcile thermal leptogenesis  
with
gravitino cosmology within minimal SUSY seesaw  
leptogenesis.
However, the mechanism proposed in \cite{hlnps} employed
neutrino Yukawa couplings much larger than suggested by
the seesaw mechanism, such that the observed neutrino masses 
were reproduced thanks to cancellations.

In this work we sketched characteristics of models,
which achieve this goal avoiding accidental cancellations by employing
the right-handed neutrino  mass matrix of  
the form~(\ref{texturem}) and
large Yukawa couplings $N_3(\ltv\cdot \vec{L})H$ with
$\lambda_3\sim 0.1$. The texture~(\ref{texturem}) allows to
control the contribution of the large neutrino Yukawa couplings in the
seesaw mechanism avoiding contribution to neutrino masses
much larger than their observed values.
We outlined how flavour models can justify the texture zeroes in eq.~(\ref{texturem}).
Assuming also $\tilde{m}_1 \sim  
10^{-3}\,{\rm eV} \ll m_{\rm sun}$
(so that thermal leptogenesis is maximally efficient),
we found that the flavour structure of the big Yukawa coupling is related to neutrino  
masses and mixings (see eq.~(\ref{venti})).
In conclusion, the models predict a detectable $\bmeg$ and
a precise correlation with $\bteg$  and $\btmg$.

We would like to finish by remarking that, although
we do not claim that our models are the only ones
that achieve the goal described above, we could not find any other solution.
Similarly, although the smallness of $\tilde m_1$ is not implied by the 
previous assumptions, 
in most of our parameter space it is  difficult to avoid it.
Future measurements of sparticle masses can clarify this issue.
In this weak sense the predictions we obtained might be a generic testable
consequence of natural realizations of the mechanism suggested in~\cite{hlnps}.

\paragraph{Acknowledgments}
We thank S.~Davidson for many discussions. A.S. thanks F.\ Feruglio.  
K.T.~is indebted to S.~Pokorski for continuous support and to  
P.~H.~Chankowski for useful comments.
K.T.~also thanks the European Centre for Theoretical Studies in
Nuclear Physics and Related Areas in Villazzano (Trento), Italy, where  
he participated in the {\em
Marie Curie} Doctoral Training Programme, for hospitality and  
stimulating atmosphere during the preparation of this work.
The work of M.~R.~is partially supported by the ESF Grant 6140, by the EC MC contract MERG-CT-2003-503626 and by the Ministry of  
Education and Research of the Republic of Estonia.
The work of K.~T.~is partially supported by the Polish State Committee  
for Scientific Research Grants 2~P03B~129~24 for years 2003-05 and  
1~P03D~014~26 for years 2004-06.


\newpage

\refstepcounter{equation} 

\appendix 
\begin{center}

\centerline{\Large\bf Addendum to}\vspace{2mm}
{\large {\bf\Large Low-scale standard  supersymmetric  leptogenesis }}\\
\vspace*{1cm}
{\bf Martti Raidal}$^1$,
{\bf Alessandro Strumia}$^2$  and {\bf Krzysztof Turzy{\'n}ski}$^{3}$\\
\vspace{0.3cm}

{\em
$^1$ National Institute of Chemical Physics and Biophysics,
Tallinn 10143, Estonia
\\
$^2$  Dipartimento di Fisica dell'Universit{\`a} di Pisa and INFN,  
Italia\\
$^3$  Institute of Theoretical Physics, Warsaw University, ul.~Ho\.za  
69, 00-681 Warsaw, Poland}
\end{center}

\setcounter{equation}{0}
\renewcommand{\theequation}{A.\arabic{equation}}
In~\cite{Raidal} we proposed a supersymmetric leptogenesis model where the two heavier right-handed neutrinos $N_{2,3}$
have Yukawa couplings naturally larger than what suggested by neutrino masses.
This allows to avoid the potential conflict 
between supersymmetry and thermal leptogenesis~\cite{DI}
by enhancing the CP-asymmetry $\varepsilon_1$ in $N_1$ decays without significantly reducing the efficiency $\eta$.

As pointed to us by S.\ Davidson,
when studying the reduction in $\eta$ we only considered the wash-out produced by
on-shell exchange of $N_{2,3}$:
off-shell exchange of $N_{2,3}$ can give an extra important contribution.

It is useful to understand from a more general point of view that off-shell 
wash-out is directly linked to the mechanism that enhances $\varepsilon_1$.
While in the standard leptogenesis scenario $\varepsilon_1$ is dominantly generated
by the dimension-5 neutrino mass operator mediated by $N_{2,3}$,
in our scenario $N_{2,3}$ mediate the dimension-7 operator $(LH) \partial^2 (LH)$,
that dominantly  contributes both to $\varepsilon_1$ and to off-shell wash-out.
It is therefore crucial to ensure that a sizable enhancement of $\varepsilon_1$ 
does not imply a sizable reduction of $\eta$.
The relevant non-renormalizable effective operators present at energies around the $N_1$ mass are
the supersymmetrized version of
\beq\Lag_{\rm eff} = c_5 \frac{(LH)^2}{2} + c_7\frac{(LH)\partial^2(LH)}{2} + \hbox{h.c.}\eeq
In our model $c_5=0$ and $c_7 = -\lambda_3^2(M_3-M_2)/M_2^2M_3^2$.
Neglecting the $c_5\cdot c_7$ interference term,
these operators  contribute to the equilibrium space-time densities of $\Delta L=2$
wash-out scatterings at temperature $T$ as
\beq \gamma_{Ns}(LH\leftrightarrow\bar L\bar H)=\gamma_{Nt}(LL\to\bar H\bar H) 
\stackrel{T\ll M_{2,3}}{\simeq} \frac{3T^{6}}{8\pi^5} |c_5|^2+
\frac{120T^{10}}{\pi^5} |c_7|^2.\eeq
We followed the notation of~\cite{gnrrsa}.
Analogous expressions hold for $\Delta L=2$ scatterings involving sparticles.
The CP asymmetry is
\beq \varepsilon_1 = \frac{3M_1}{8\pi}{\rm Im}\, c_5 - \frac{3M_1^3 }{16\pi}{\rm Im}\,c_7\eeq
in agreement with eq.~(\ref{eq:eps1}) of~\cite{Raidal}.
The above operator approximation has a general validity and shows that
a too large off-shell washout would exclude the whole scenario, not only our model.
One can analytically check that in most of the parameter space
off-shell washout is small enough.

We here present a full numerical computation of thermal leptogenesis
that includes all contributions to wash-out
as predicted by our supersymmetric model.
 Notice that the generic expressions  for the $\Delta L=2$ wash-out
 scattering rates~\cite{Plumacher}
 are not sufficiently accurate in our very specific context.  
 Indeed these expressions are obtained by modifying the propagators
 of the mass-eigenstate right-handed neutrinos $N_{i}$  and sneutrinos $\tilde{N}_i$ ($i=\{1,2,3\}$)
  by taking into account the decay widths $\Gamma_i$ according to
 the usual prescription: 
 \beq \Pi_{ii}(p) = \frac{1}{p^2-M_i^2 + i \Gamma_i M_i}.\eeq
We see that $\Pi_{ii}(0)$ gets modified, spoiling the delicate cancellation that in our model suppresses the dimension-5 operator.
 To correctly reproduce this cancellation one has to take into account  that also
 the  
 off-diagonal $\Pi_{ij}$ propagators develop an imaginary part.

\begin{figure}
\begin{center}
$$\includegraphics[height=7cm]{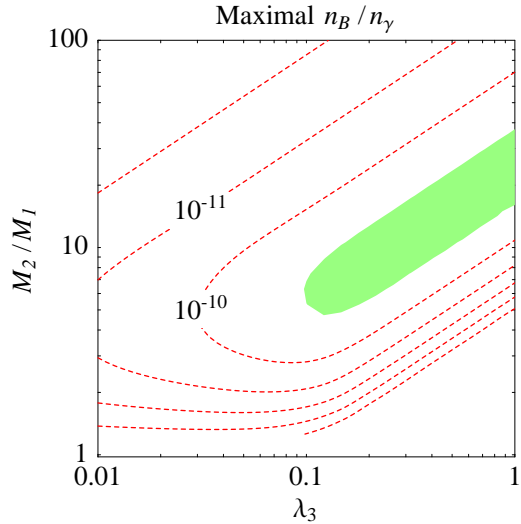}$$\vspace{-10mm}
\caption[1]{\em \label{fig:new}
Fig.~2b of~{\rm\cite{Raidal}} modified taking into account all
contributions to $\Delta L=2$ scatterings.
 We assumed   $M_1 = 10^6\gev$, $M_3/M_2 = 2$, $\tilde{m}_1 = 10^{-3}\,\hbox{\rm  
eV}$. }
\end{center}
\end{figure}

Fig.~\ref{fig:new} shows our full result, that can be directly compared with fig.~2b in the paper~\cite{Raidal}, 
where the off-shell effect was not included.
The new off-shell washout is more important than on-shell washout
only when $\lambda_3\sim M_2/M_1\sim 1$.
As a result, the region where leptogenesis can be successful
survives  and is shaded in green in fig.~\ref{fig:new}, getting somewhat reduced.
The constraint on the parameter that controls lepton-flavour violating signals
remains unaltered: $\lambda_3\circa{>}0.1$.

\end{document}